\documentclass[prd,superscriptaddress,showpacs,nofootinbib,amsmath,amssymb]{revtex4} 

\usepackage{bm}
\usepackage{amsfonts}
\usepackage{latexsym}
\usepackage[latin1]{inputenc}
\usepackage{graphicx}
\usepackage{amsmath}
\usepackage{rotating}
\usepackage{epsfig}

\newcommand{\ep}{\epsilon}
\newcommand{\om}{\omega}
\newcommand{\Om}{\Omega}
\newcommand{\al}{\alpha}
\newcommand{\la}{\lambda}
\newcommand{\YY}{Y_{lm}}
\newcommand{\Yph}{\frac{\partial Y_{lm}}{\partial \phi}}
\newcommand{\Yth}{\frac{\partial Y_{lm}}{\partial \theta}}
\newcommand{\I}{\mbox{\rm i}}
\def \nn  {\nonumber}

\def \eps{\epsilon}

\def\jnl@style{\it}
\def\aaref@jnl#1{{\jnl@style#1}}

\def\aaref@jnl#1{{\jnl@style#1}}

\def\aj{\aaref@jnl{AJ}}                   
\def\apj{\aaref@jnl{ApJ}}                 
\def\apjl{\aaref@jnl{ApJ}}                
\def\apjs{\aaref@jnl{ApJS}}               
\def\apss{\aaref@jnl{Ap\&SS}}             
\def\aap{\aaref@jnl{A\&A}}                
\def\aapr{\aaref@jnl{A\&A~Rev.}}          
\def\aaps{\aaref@jnl{A\&AS}}              
\def\mnras{\aaref@jnl{MNRAS}}             
\def\prd{\aaref@jnl{Phys.~Rev.~D}}        
\def\prl{\aaref@jnl{Phys.~Rev.~Lett.}}    
\def\qjras{\aaref@jnl{QJRAS}}             
\def\skytel{\aaref@jnl{S\&T}}             
\def\ssr{\aaref@jnl{Space~Sci.~Rev.}}     
\def\zap{\aaref@jnl{ZAp}}                 
\def\nat{\aaref@jnl{Nature}}              
\def\aplett{\aaref@jnl{Astrophys.~Lett.}} 
\def\apspr{\aaref@jnl{Astrophys.~Space~Phys.~Res.}} 
\def\physrep{\aaref@jnl{Phys.~Rep.}}      
\def\physscr{\aaref@jnl{Phys.~Scr}}       

\begin{document}

\title[Differential Rotation]{Nonradial oscillations of slowly and differentially rotating compact stars}


\author{Adamantios Stavridis}
\author{Andrea Passamonti}
\author{Kostas Kokkotas}
\affiliation{Department of Physics, Aristotle University of Thessaloniki 54124, Greece}

\date{\today}

\begin{abstract}
The equations describing nonradial adiabatic oscillations of
differentially rotating relativistic stars are derived in relativistic
slow rotation approximation.  The differentially rotating
configuration is described by a perturbative version of the 
relativistic j-constant rotation law.
Focusing on the oscillation properties of the stellar fluid, the
adiabatic nonradial perturbations are studied in the Cowling
approximation with a system of five partial differential equations. In
these equations, differential rotation introduces new coupling terms
between the perturbative quantites with respect to the uniformly
rotating stars.  In particular, we investigate the axisymmetric and
barotropic oscillations and compare their spectral properties with
those obtained in nonlinear hydrodynamical studies.  The perturbative
description of the differentially rotating background and the
oscillation spectrum agree within a few percent with those of the
nonlinear studies.
\end{abstract}


\maketitle

\section{Introduction}

Differential rotation can appear in many phases of the stellar
evolution of a neutron star, such as in protoneutron
stars~\cite{2002A&A...393..523D,2004ApJ...600..834O}, in the massive
remnant of binary neutron star
mergers~\cite{1994ApJ...432..242R,Shibata:1999wm}, or as a results of
stellar oscillations (r-modes) that may drive the star into
differential rotation via nonliner
effects~\cite{2000ApJ...531L.139R,2001MNRAS.324..917L,2001PhRvL..86.1148S,2004PhRvD..69h4001S}.
The differentially rotating phase can last at most some seconds
depending on the dissipative mechanism that drive the star to uniform
rotation, such as magnetic bracking~\cite{2000ApJ...544..397S,
2003ApJ...599.1272C} and turbulent motion~\cite{1977ApJ...217..244H,
liu:044009}.

However, this very short period can
appear during the most violent phases of neutron star's life, as in
the case of a core collapse or binary merging. This is exactly the
time when we expect to get the strongest emission in gravitational
waves.
 Since the ground based
detectors reached sensitivities which allow the detection of
gravitational wave signals from oscillating or unstable neutron stars
the exact frequencies of the emitted waves are urgently needed.
The study of linear pulsations and stability of differentially
rotating stellar models started more than three decades ago both in
Newtonian and relativistic gravity
\cite{1972ApJS...24..319S,1972ApJS...24..343S,1973ApJ...179..289S,1975ApJ...197..745S}.
The spectrum of differentially rotating stars in Newtonian theory has
been studied in the frequency domain by~\cite{1977ApJ...217..151H}.

Recently, the effects of differential rotation on the dynamical and
secular instabilities of the r and f-modes gained more attention.  In
Newtonian theory, the r-mode spectrum has been studied
in~\cite{2001PhRvD..64b4003K,Abramowicz:2000su,Rezzolla:2001qt}, while
the f-mode and the secular stability limits have been investigated
in~\cite{2002ApJ...568L..41Y}.
In stars that rotate with a high degree of differential rotation
($\Omega_c/\Omega_s\approx 10$), an $m=2$ dynamical instability can
appear even for considerably low rotation rates ($T/|W|\sim
O(10^{-2})$) as suggested
in~\cite{2002MNRAS.334L..27S,2005PhRvD..71b4014S}.  Note that $T$ is
the rotational kinetic energy, $W$ the gravitational potential energy,
$\Omega_c$ and $\Omega_s$ the angular velocity at the center and at
the equatorial surface, respectively. In addition, an $m=1$ 
dynamical instability has been identified for high degrees of
differential rotation and soft equations of state
\cite{2001ApJ...550L.193C,2003ApJ...595..352S}.  More recently, it has
been reported the discovery of $m=1$ and $m=2$ dynamical instability
even for stiff equations of state \cite{2006ApJ...651.1068O}.
Studies based on linear analysis
\cite{2005ApJ...618L..37W,2006MNRAS.368.1429S} suggest that low
$T/|W|$ instabilities might be triggered when the corotation points of
the unstable modes fall within the differentially rotating structure
of the star.
It is then evident that differentially rotating compact
stars got a renewed interest and there are several recent 
fully general relativistic studies related to the issue. 
For example,  slowly and
differentially rotating magnetized neutron stars have been studied in
\cite{Etienne:15,Duez:2006qe}, the collapse and black hole formation
of magnetized and differentially rotating neutron stars in
\cite{Stephens:2006cn} and the  frequencies for nonlinear
axisymmetric pulsations of differentially rotating stars in the
Cowling and conformal flatness
approximations in ~\cite{Stergioulas:2003ep,Dimmelmeier:2005zk}.

As a further contribution to the direction of understanding stellar
differential rotation we present here the equations describing a
slowly and differentially rotating relativistic star. In our study we
use a perturbative analysis and keep terms up to first order with
respect to stellar rotation.  In this first perturbative approach, we
neglect all the spacetime perturbations i.e. we use the so called
Cowling approximation since we want to focus on the behavior of the
stellar fluid.  The slow rotation approximation has been extensively
used in the study of stellar perturbations both in Newtonian and in
general relativistic approach
\cite{Kojima:1992kj,1999ApJ...521..764L,2001MNRAS.328..678R,2005IJMPD..14..543S}.
Here we study the spectrum of axisymmetric perturbations in order to
test the accuracy of our approximation technique against published
results by nonlinear numerical codes
\cite{Stergioulas:2003ep,Dimmelmeier:2005zk}.  The next steps will be
to estimate the oscillation spectrum of the non-axisymmetric
perturbations and furthermore to proceed in studying the low $T/|W|$
dynamical instability.

The structure of the paper is as follows. In section~\ref{sec:Eqconf}
we construct the differentially rotating stellar models, and test
their accuracy against the equilibrium configurations obtained with
non-linear codes. In section III, we derive the perturbation equations
in the general case of non-barotropic and non-axisymmetric
perturbations, which are numerically solved in
section~\ref{Sec-Num-Res} for the axisymmetric and barotropic case.
In the section \ref{conclusions}, conclusions are drawn and the future
applications are proposed.  The appendix is organized in four 
sections. Section~\ref{Rot-law-Int} is dedicated to the analytical
expressions of the harmonic components of the j-constant rotation law,
while all the coefficients of the perturbed conservation equations are
given in section~\ref{Perturb-terms}.  The operators that couple
perturbations with different harmonic indices are introduced in
section~\ref{Integrals}. Finally, in the last
section~\ref{sec:rot-law-isotr} we discuss the j-constant rotation law
in isotropic coordinates.

Throughout the paper we use geometrical units $c=G=1$. We use a prime (~$'$~) to
denote derivatives with respect to the radial coordinate $r$ and the
overdot (~$\dot{}$~) for derivatives with respect to the time coordinate
$t$.

\section{Equilibrium configuration} \label{sec:Eqconf}

The axially symmetric spacetime of a slowly and differentially
rotating star can be described by the following line element:
\begin{align}
ds^2 = - e^{2\nu} dt^2 + e^{2\la} dr^2 & - 2 \, \omega \, r^2 \sin ^2
    \theta \, dt \, d \phi  + r^2 \left( d \theta ^2 + \sin^2
    \theta d \phi ^2 \right) \, , \label{ds-equil}
\end{align}
where $\nu$, $\la$ are scalar fields that depend on the radial
coordinate $r$. The metric function $\omega$, which describes the
dragging of the reference frames due to the stellar rotation, 
is a function of the both $r$ and $\theta$ coordinates.  In this paper, we assume that the
stellar interior is described by the perfect fluid energy-momentum
tensor:
\begin{equation}
T^{\alpha \beta} = \left( \epsilon + p \right) u^{\alpha} u^{\beta} +
p g^{\alpha \beta} \, , 
\end{equation}
where $\ep$ and $p$ denote the total energy density and the pressure
respectively, and $u^{\alpha}$ is the fluid velocity 
\begin{equation}
u ^{\al} = \left( e^{-\nu}, 0, 0, \Omega e^{-\nu} \right) \, ,
\label{bg-vel} \end{equation}
where $\Omega = \Omega \left(r , \theta \right)$ is the angular
velocity of the star as measured by an observer at infinity.

By providing an equation of state (EoS) $p = p\left( \ep \right)$, a
rotating equilibrium configuration can be determined by solving the
Tolman-Oppenheimer-Volkoff~(TOV) equations together with the following equation for the metric function   $\omega$:
\begin{widetext}
\begin{eqnarray}
\omega^{''} - \left[ 4 \pi \left( \ep + p \right) r e^{2\la} 
                - \frac{4}{r} \right] \omega^{'} 
                - \left[ 16 \pi \left( \ep + p \right) + \frac{l\left(l+1\right) - 2 }{r^2} 
                  \right] e^{2\la} \omega
                = - 16 \pi \left( \ep + p \right) e^{2\la} \Omega \, . 
\label{drag-eq}
\end{eqnarray}
\end{widetext}


\subsection{Rotation Law} \label{Sec-Rot-law}

In general the differential profile of the angular velocity $\Omega$ in not known. Still approximate
analytic differential rotating laws have been adopted and used both for Newtonian and General Relativistic configurations
(see~\cite{Stergioulas:2003yp} and reference therein).  Among these
laws, the j-constant rotation
law~\cite{Komatsu1989MNRASA,Komatsu1989MNRASB} guarantees Rayleigh's
stability against axisymmetric perturbations for inviscid fluids, and
seems to describe quite well not only typically differentially rotating stars
but even the remnant rotational configurations that arise from
the merging of hypermassive binary systems~\cite{Shibata:1999wm}. 

The relativistic j-constant rotation law can be perturbatively
expanded as in \cite{Passamonti:2005cz}:
\begin{equation}
\Omega(r,\theta) = \frac{A^2\Omega^{}_{c} + e^{-2\nu} \omega(r,\theta) r^2 \sin^2\theta\,}
{A^2 + e^{-2\nu} r^2 \sin^2\theta} \, , \label{j-cons}
\end{equation}
where $\Omega_c$ denotes the angular velocity at the rotation axis,
 while $A$ is the parameter that describes the degree of differential
 rotation of the star. For high values of $A$ (e.g. $A\sim 500$ km), the j-constant rotation
 law~(\ref{j-cons}) tends to a uniformly rotating configuration, where
 $\Omega \rightarrow \Omega_c$.

In the slow rotation approximation, rotation is treated as a
perturbation of a spherically symmetric background spacetime. Therefore, the
tensor harmonic expansion enables us to separate the angular dependence
of any perturbative fields from the temporal and radial parts.
In this way the metric function~$\omega$ and the angular velocity $\Omega$ of the
fluid can then be written as follows:
\begin{eqnarray}
\omega(r,\theta) & = & - \sum _{l} \omega_{l} (r) \frac{1}{\sin \theta} 
\frac{\partial P_{l}}{\partial \theta} \, , 
\label{om-exp}\\ 
\Omega(r,\theta) & = & - \sum _{l} \Omega_{l} (r) \frac{1}{\sin \theta} 
\frac{\partial P_{l}}{ \partial \theta} \, , 
\label{Om-exp}
\end{eqnarray} 
where $P_{l}(\cos\theta)$ are the Legendre's polynomials.  The $
\Omega_{l}$ and $\omega_{l}$ components in
equations~(\ref{om-exp})-(\ref{Om-exp}) are then determined via
the orthogonality relations of the form,
\begin{equation} 
f_{l} \left( r \right) = - \frac{1}{4\pi} \frac{2 l +1 }{l\left( l+1\right)} 
                               \int_{S^2} \, d \tilde{\Omega} \, 
                               f \left(r , \theta \right) 
                               \sin \theta \, \frac{ \partial P_{l}}{\partial \theta} 
                                \, ,  \label{Om_lm}
\end{equation}
where  $f=f(r,\theta)$ is a scalar field.  In
equation~(\ref{Om_lm}), $d\tilde \Omega$ denotes the volume element of
the 2-sphere~$S^2$.  It is worth mentioning that for a uniformly rotating
star, the only non vanishing components of
equations~(\ref{om-exp})-(\ref{Om-exp}) are the ones for $l=1$
\cite{Hartle:1967ha}.  In this case, $\omega$ depend only on the radial coordinate $r$ 
and $\Omega$ is of course constant.
For differentially rotating stars, which are described by the 
j-constant rotation law, the components $\Omega_l$ can be
determined by direct substitution of equation~(\ref{j-cons})
in~(\ref{Om_lm}). Then the resulting expression can be split in two
parts: 
\begin{equation}
\Omega_{l} = \mathcal{I}_{l}^{N} + \mathcal{I}_{l}^{R} \, ,  \label{Om-tot}
\end{equation}
where $\mathcal{I}_{l}^{N}$ is the ``nearly Newtonian'' part  and 
$\mathcal{I}_{l}^{R}$  is the ``relativistic'' correction.  
The definition of these two terms is given by the following expressions:
\begin{align}
& \mathcal{I}_{l}^{N}   \equiv  - \frac{2l+1}{l\left( l+1 \right)}  
                             \frac{\Omega^{}_{c}}{4\pi}
                             \int_{S^2}  d \tilde{\Omega} \,
                             \frac{ A^2 \sin \theta}
		             {A^2 + e^{-2\nu} r^2 \sin^2\theta}
                              \frac{\partial P_{l}}{\partial \theta}  
                               \, , \label{In-def}\\ 
& \mathcal{I}_{l}^{R}   \equiv  - \frac{2l+1}{l\left( l+1 \right)}  
                             \frac{e^{-2\nu} r^2}{ 4 \pi }  
                             \int_{S^2}  d \tilde{\Omega} \,  
                             \frac{ \sin^3\theta\, \omega(r,\theta)}
                               {A^2 + e^{-2\nu} r^2 \sin^2\theta} 
                               \frac{\partial P_{l}}{\partial \theta} 
                              \, .  \label{Ir-def}
\end{align}
The parity of the  j-constant rotation law implies that the non-vanishing 
components in~(\ref{In-def})-(\ref{Ir-def}) are the ones with odd $l$.
The integral  $\mathcal{I}_{l}^{N}$ for $l=1$ and $l=3$ has been already
calculated in~\cite{Passamonti:2005cz}, and is analytically written in
appendix~\ref{Rot-law-Int}. Moreover, the integration of the
``relativistic'' term $\mathcal{I}_{l}^{R}$ requires an additional expansion of the
metric variable $\omega$. 
Therefore, by inserting equation (\ref{om-exp})
into (\ref{Ir-def}) one gets
\begin{equation}
\mathcal{I}_{l}^{R}  =  \frac{2l+1}{l\left( l+1 \right)}  
                        \frac{e^{-2\nu} r^2}{ 4 \pi }  
			\sum_{l^{'}} \omega_{l^{'}} \left( r \right)  
                         \mathcal{J}^{~l^{'}} _{l} \, ,  \label{Ir-exp}
\end{equation}
where $\mathcal{J}^{l^{'} }_{l} $ are given by the following integral relation
\begin{equation}
\mathcal{J}^{l^{'} }_{l} \equiv \int_{S^2} d \tilde{\Omega} \, \frac{
                                \sin^2 \theta} {A^2 + e^{-2\nu} r^2
                                \sin^2\theta} \frac{\partial
                                P_{l}}{\partial \theta} \,
                                \frac{\partial P_{l'}}{\partial
                                \theta} \, .   \label{Jlm-def}
\end{equation}
The detailed forms of $\mathcal{J}^{1 }_{1}$, $\mathcal{J}^{3}_{1}$ and 
$\mathcal{J}^{3}_{3}$ are shown in appendix~\ref{Rot-law-Int}. 
Our choice to retain only the $l=1$ and $l=3$ terms for the expression~(\ref{Jlm-def}) 
will be justified later in section~\ref{Sec-Num-Res}, were we show that 
the frequencies of the nonradial oscillations that we calculate using this 
approach are in very good agreement with the 
ones derived using the full nonlinear evolution of the fluid background
\cite{Stergioulas:2003ep,Dimmelmeier:2005zk}.

\subsection{Frame dragging in differential rotation}

The various components $\omega_l$ of the series expansion (\ref{om-exp}) which 
provides $\omega$ will be determined from the solution of equation 
(\ref{drag-eq}) for the frame dragging. In practice, one substitutes 
equations (\ref{om-exp}) and (\ref{Om-exp}) into
equation~(\ref{drag-eq}), and collects the terms of the series with
the same index $l$.
However, the term $\mathcal{I}_{l}^{R}$ describing the relativistic correction, 
equation (\ref{Ir-exp}), introduces a coupling between the
different harmonics, which is due to the presence of metric function $\omega$ in
equation~(\ref{j-cons}). As a result the terms $\omega_l$ will be determined
as solutions of a system of coupled ordinary differential equations.
In order to simplify the presentation of this system of differential equations
we will write it schematically as:
\begin{equation}
\mathcal{L} \left( \omega \right) = \mathcal{S} \left( \Omega \right)  \, , \label{sc-drag}
\end{equation}
where $\mathcal{L} ( \cdot )$ represents the linear differential
operator of the left hand side of equation~(\ref{drag-eq}), while
$\mathcal{S} (\cdot)$ denotes the linear operator of the source.
When equation (\ref{sc-drag}) is expanded into vector harmonics, it reduces 
to the following set of equations
\begin{equation}
\mathcal{L} \left( \omega _{l} \right) = \mathcal{S} \left( \mathcal{I}_{l}^{N} \right)   
           + \frac{2l+1}{l\left( l+1 \right)} \frac{e^{-2\nu} r^2}{ 4 \pi }  
              \sum_{l^{'} } 
            \mathcal{S} \left( \omega_{l^{'} } \mathcal{J}_{l}^{~l^{'}} \right) \, , \label{sc-drag-exp}
\end{equation}
which applies to any value of $l$. The first two non-vanishing terms 
 $\omega_1$ and $\omega_3$ obey the following coupled system of 
equations:
\begin{widetext}
\begin{eqnarray}
\mathcal{L} \left( \, \omega _{1} \, \right) - \frac{3}{ 8 \pi }  r^2 e^{-2\nu}  
                   \mathcal{S} \left( \, \omega_{1} \, \mathcal{J}_{1}^{1} \, \right) 
             & = & \mathcal{S} \left( \, \mathcal{I}_{1}^{N} \, \right)   
                    + \frac{3}{ 8 \pi }  r^2 e^{-2\nu}  \mathcal{S} \left( \, \omega_{3} \, \mathcal{J}_{1}^{3} \, \right) 
                    \, , \label{drag-eq10} \\
\mathcal{L} \left( \, \omega _{3} \, \right) - \frac{7}{48\pi}  r^2 e^{-2\nu} 
                   \mathcal{S} \left( \, \omega_{3} \, \mathcal{J}_{3}^{3} \, \right) 
             & = & \mathcal{S} \left( \, \mathcal{I}_{3}^{N} \, \right)   
                   + \frac{7}{48\pi}  r^2 e^{-2\nu} \mathcal{S} \left( \, \omega_{1} \, \mathcal{J}_{3}^{1} \, \right) 
                   \, , \label{drag-eq30} 
\end{eqnarray} 
\end{widetext}
which  together with a set of appropriate boundary conditions at
the stellar center and at infinity form a boundary value problem.  
Actually, the demand for regularity of the solutions at the stellar center 
and their decaying asymptotic behavior suggest the following 
approximate relations~\cite{Hartle:1970ha}:
\begin{eqnarray}
\omega_l & \sim & r^{l-1} \, , \quad \, \, \, \, {\rm for} \quad r \to 0 \, , \label{bc:origin}\\   
\omega_l & \sim & r^{-l-2} \, , \quad {\rm for} \quad r \to \infty \, .  \label{bc:infinity}
\end{eqnarray}
The numerical method used for the solution of the above system of equations
is described in the next paragraphs.

\subsection{Stellar Models}

In order to test the accuracy of the perturbative treatment of stellar 
differential rotation, which has been introduced in this paper, we will compare our 
results with those derived from the numerical solution of the nonlinear Einstein equations.
Thus we will study a set of stellar models, the so-called B
models, which have been already adopted in 
\cite{Stergioulas:2003ep, Dimmelmeier:2005zk}.  These models
describe barotropic and differentially rotating stars, where the
rotating pattern is described by the j-constant rotation law.  For simplicity,
a relativistic barotropic EoS of the form
\begin{eqnarray}
p & = & K \rho ^{\Gamma} \, ,\\
\epsilon & = & \rho + \frac{p}{\Gamma -1} \, ,
\end{eqnarray}
has been adopted, where $\rho$ is the rest mass density and $K$ and $\Gamma$ the
polytropic parameters.  
%
\begin{table}[t]
\begin{center}  
\begin{ruledtabular}
\begin{tabular}{*{4}{c}}
  Model  &  $T_c$ (ms) & $\Om_c$ ($\times 10 ^{-2}$)$~{\rm km}^{-1}$ & 
  $\Om_e$ ($\times 10^{-2}$)$~{\rm km}^{-1}$ \\
\hline 
  B1  &  1.719  & 1.218 &  0.435  \\
  B2  &  1.204  & 1.740 &  0.621  \\
  B3  &  0.970  & 2.160 &  0.771  \\
  B6  &  0.657  & 3.189 &  1.139  \\
  B9  &  0.496  & 4.219 &  1.507  \\
\end{tabular}
\end{ruledtabular}
\caption{\label{tab:Models} Properties of the background models used
in the paper. The quantities $T_c$ and $\Om_c$ are respectively the
period and the angular velocity at the rotational axis, while $\Om_e$
represents the angular velocity at the equator. All stellar models 
have mass $M=1.40~M_{\odot}$ and radius $R = 14.151~{\rm km}$.}
\end{center}
\end{table}

For a given EoS, an equilibrium configuration
depends on three parameters: the central density $\rho_c$, the stellar
spin at the rotation axis~$\Omega_c$ and the
parameter $A$ describing the degree of differential rotation. 
For the B models the polytropic parameters  get the values
$K = 217.858~{\rm km^2}$ and $\Gamma = 2$ while the central rest mass density
is fixed to the value ${\rm \rho_c} = 5.87 \times 10^{-4}~{\rm km}^{-2}$. 
The solution of the TOV equations provide a stellar model with mass $M=1.40~M_{\odot}$
and radius $R = 14.151~{\rm km}$.
The angular velocity at the rotation axis, $\Omega_c$, for some of the
B1-B9 models is listed in table~\ref{tab:Models}.  In order to compare
our results with those of nonlinear studies special care has been
taken for the interpretation of the parameter $A$, due to the
different coordinate systems that are commonly used in non-linear
studies and our perturbative approximation.  This issue is explained
in detail in appendix~\ref{sec:rot-law-isotr}, where we show that $A$
should be set to the value of the isotropic stellar radius.

\begin{figure}[!t]
\begin{center}
\includegraphics[width=85mm,height=80mm]{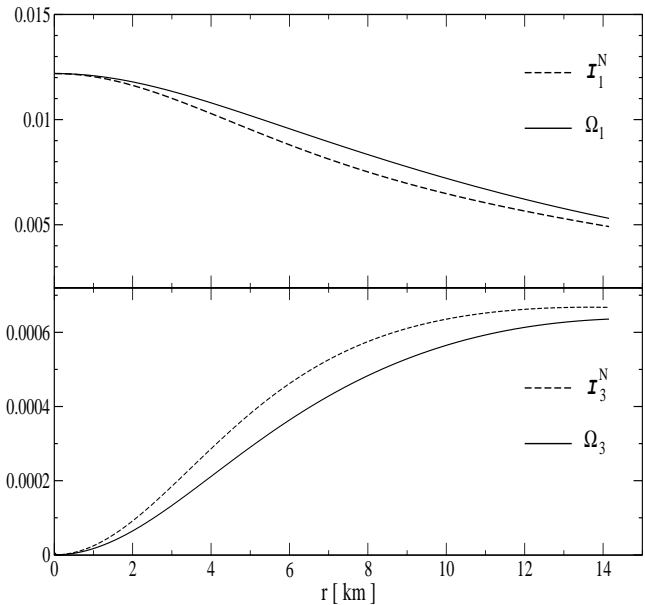}
\caption{The harmonic components $\Omega_1$ (\emph{upper
panel}) and $\Omega_3$ (\emph{lower panel}) 
of the angular velocity $\Om$ of the B1 stellar model are plotted, 
where $\Omega_{c}= 1.22
10^{-2}~\rm{km}^{-1}$ and $A=12~\rm{km}$. 
The \emph{dashed lines} display the contribution of the ``nearly Newtonian term''
given by equation (\ref{In-def}).
\label{fig:Om-lm}}
\end{center}
\end{figure}

\begin{figure}[t]
\begin{center}
\includegraphics[width=85mm,height=80mm]{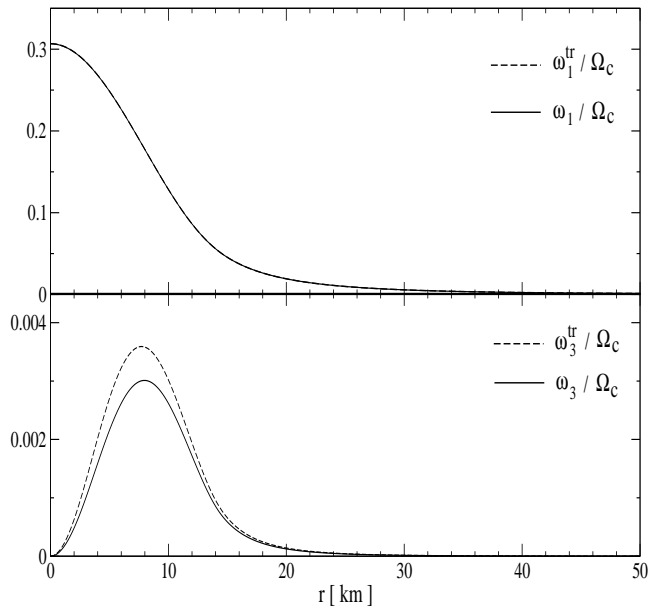} 
\caption{The harmonic components $\omega_1$ (\emph{upper
panel}) and $\omega_3$ (\emph{lower panel}) 
of the angular velocity $\omega$ of the B1 stellar model are plotted. 
The \emph{dashed lines} denote the trial solutions of
equation~(\ref{drag-eq10}), which have been determined without the
coupling terms $\mathcal{J}_{3}^{1}$ and $\mathcal{J}_{1}^{3}$.  The
\emph{solid lines} are instead the final solutions. 
\label{fig:drag-lm}}
\end{center}
\end{figure}

As we mentioned earlier the two components $\omega_1$ and $\omega_3$
of the metric function $\omega$ will be the solutions of the boundary
value problem described by the two differential equations
(\ref{drag-eq10})-(\ref{drag-eq30}) together with the boundary
conditions (\ref{bc:origin})-(\ref{bc:infinity}).  By discretizing
equations (\ref{drag-eq10})-(\ref{drag-eq30}) using a second order
scheme one can construct a tridiagonal linear system. This system can
be solved using a standard $\mathbf{LU}$ decomposition where the
boundary conditions~(\ref{bc:origin}) and~(\ref{bc:infinity}) are
appropriately implemented. The presence of coupling terms on the right
hand side of the equations calls for an iterative treatment.
As a first step a trial solution $\omega_l^{\rm{tr}}$
of~(\ref{drag-eq10})-(\ref{drag-eq30}) can be determined by neglecting
the coupling source terms, i.e. $\mathcal{J}_1^3 = \mathcal{J}_3^1 =
0$. This trial solution is then substituted only into the source terms
and the system of equations~(\ref{drag-eq10})-(\ref{drag-eq30}) is
solved. This procedure is repeated until the solution converges. In
practice, only three or four iterations are needed.

\begin{figure}[!t]
\begin{center}
\includegraphics[width=85mm,height=80mm]{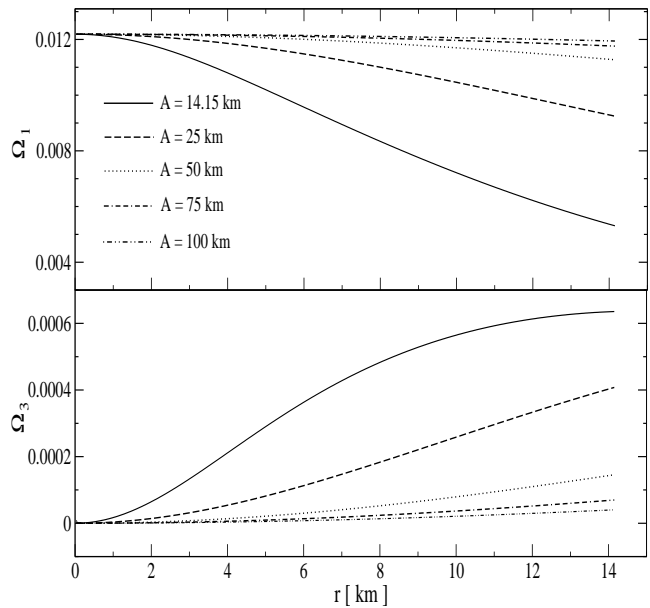}
\caption{The dependence of the two components of the angular velocity $\Om$ 
on the differential parameter $A$ is shown in this figure for the stellar model 
B1.  
\label{fig:A-Om-lm}}
\end{center}
\end{figure}

\begin{figure}[t]
\begin{center}
\includegraphics[width=85mm,height=80mm]{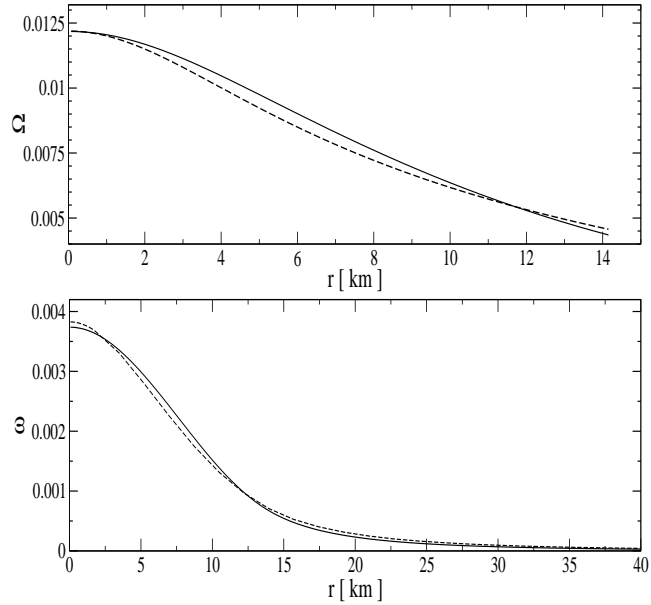}
\caption{The profiles of $\Omega$ and $\omega$ derived from the perturbative 
treatment and the nonlinear RNS code
compared in this figure. 
The \emph{solid lines} represent the perturbative solutions and
\emph{dashed lines} the RNS quantities. In the \emph{upper panel}
the estimations of the angular velocity $\Om$ are plotted, while 
in the \emph{lower panel} the variable $\omega$ is compared to the shift 
component $-\beta^{\phi}$, which have been determined with the RNS code.
\label{fig:Om-om-tot}}
\end{center}
\end{figure}

In figures~\ref{fig:Om-lm}-\ref{fig:Om-om-tot}, we show the behavior
of the angular velocity $\Omega$ and the metric function $\omega$ for
the slowly rotating model B1, see table~\ref{tab:Models}.
In figure \ref{fig:Om-lm} it is noticeable the small contribution of
the ``relativistic corrections'' $\mathcal{I}_{R}^{l}$ in the
estimation the star's angular velocity. The two components of the
metric function i.e. $\omega_1$ and $\omega_3$ are plotted in
figure~\ref{fig:drag-lm}, together with initial trial solutions
$\omega_l^{\rm{tr}}$. It is worth noticing an imperceptible correction
to the initial trial solution for $l=1$, and the small value of
$\omega_3$ with respect to $\omega_1$, which suggests that terms of
the perturbative series~(\ref{om-exp})-(\ref{Om-exp}) with higher $l$ will not
affect the results.
The dependence of $\Omega_1$ and $\Omega_3$ on the the differential
rotation parameter $A$ is illustrated in
figure~\ref{fig:A-Om-lm}. There one can notice that for increasing
values of~$A$ the $\Omega_{1}$ reaches its uniform rotation value
$\Omega_c$, while $\Omega_{3}$ decreases proportionally as expected.

In order to estimate  the accuracy of our approximate solutions for 
$\Omega$ and $\omega$, we have done a comparison with the profiles 
derived by the nonlinear numerical code RNS
~\cite{1995ApJ...444..306S}.  The results from RNS and the perturbative
solutions for the B1-model are compared in figure~\ref{fig:Om-om-tot} on
the equatorial plane.
The maximum deviation between the two estimations is around
$r=6~\rm{km}$ where the relative error is about $5.5$ \% .
The accuracy could be further improved by taking into account the
$l=5$ contributions. However, the nonradial perturbation equations
would become considerably more complicated due to the presence of
extra coupling terms.  As we show in the next section the
approximation used here allows the estimation of the nonradial
oscillations frequencies with an error smaller than 10 percent even
for very rapidly rotating models, see table~\ref{tab:Modes}.
%

\section{Perturbation Equations \label{sec:pert-eqs}}

In this section, we derive the perturbation equations which describe
the nonradial oscillations of a slowly and differentially rotating
star.  The rotation, as described in the previous section, is treated 
perturbatively up to the first order in $\Omega$~\cite{Ruoff:2001fq}.  
In this work we focus on the spectral properties of the fluid modes,
therefore, we neglect all metric perturbations, and we use 
the so-called ``Cowling approximation''.  
For slowly rotating stars, this method enables us to
determine the frequencies of nonradial pulsations with an accuracy to
better than 15 \% with respect to the non approximated
approach. In addition, the accuracy improves for modes with higher
harmonic index $l$ and mode order $n$~\cite{1997MNRAS.289..117Y}.

We will consider adiabatic oscillations, where the pressure
and density perturbations are related by the following expression:
\begin{equation}
\label{Bar_rel}
\delta p = \frac{\Gamma_1 p} {p + \eps} \delta \eps 
+ p' \xi^r \left( \frac{\Gamma_1}{\Gamma} - 1 \right) \, ,
\end{equation}
where $\xi^r$ is
the radial component of the fluid displacement vector $\xi^{\mu}$ and
$\Gamma$ and $\Gamma_1$ represent the adiabatic index of the background 
and perturbed configurations respectively. The
adiabatic index $\Gamma$ is defined by the following relation:
\begin{equation}
\Gamma = \frac{ p + \eps}{\eps } \frac{dp}{d\eps} \, ,
\end{equation}
and sound speed by
\begin{equation}
c_s^2 = \frac{\Gamma_1}{\Gamma} \frac{d p}{d\eps} \, .
\end{equation} 
In Cowling approximation, the displacement vector $\xi^\mu$ is related to the 4-velocity
perturbation by
\begin{equation}
\delta u_{\mu} = g_{\mu\nu} u^{\lambda} \frac{\partial \xi^{\nu}}{\partial x^{\lambda} } \, .  \label{u-xi-rel} 
\end{equation}
The $r$ component of this equation provides an evolution equation for
$\xi^r$:
\begin{equation}
\label{xi_r_eqn}
(\partial_t + \Omega \partial_{\phi} ) \xi^r = e^{-2\lambda+\nu} \delta u_r  \, .
\end{equation}

The three scalar quantities describing the perturbations of the total energy 
density, pressure and the $r$ component of the
Lagrangian displacement can be expanded in spherical harmonics,  
\begin{eqnarray}
\delta \epsilon & = & \delta \eps _{lm} \YY \, , \\
\delta p        & = &  \delta p _{lm} \YY \, , \\ 
\xi^r           & = & \left[ \nu' \left( 1 - \frac{\Gamma_1}{\Gamma} 
                      \right) \right]^{-1} \xi^{lm} \YY \, . \label{xir-exp} 
\end{eqnarray}
The perturbation equations can be considerably simplified if one uses
a new function $H$, which is related to the energy density and
pressure perturbations by the following relations
\begin{eqnarray} 
\delta p _{lm} & = & H_{lm} (p + \eps) \, , \\ \delta \eps_{lm} & = &
\frac{(p+\eps)^2}{p \Gamma_1} \left( H_{lm} - \xi_{lm} \right) \, .
\end{eqnarray}
For barotropic fluids, $H$ becomes the perturbation of the
relativistic enthalpy~\cite{Ipser:1991ip}.
Velocity perturbations are also expanded into vector harmonic expansion 
and have the form:
\begin{widetext}
\begin{eqnarray}
\delta u_r & = & - e^{\nu} \sum _{lm} u_{1,\, lm} \YY \, , \\ \delta
u_{\theta} & = & - e^{\nu} \sum _{lm} \left( u_{2,\, lm} \Yth - u_{3,
\, lm} \Yph\frac{1}{\sin \theta} \right)\, , \\ \delta u_{\phi} & = &
- e^{\nu} \sum _{lm} \left( u_{2,\, lm} \Yph + u_{3, \, lm} \sin
\theta \Yth \right) \, , \\ \delta u_t & = & - \Om \left( r,\theta
\right) \delta u_{\phi} = - e^{\nu} \sum _{l'm'} \sum _{lm} \left(
u_{2,\, lm} \Yph + u_{3, \, lm} \sin \theta \Yth \right) \Omega_{l'm'}
(r) \frac{1}{\sin \theta} \frac{\partial P_{l'}}{\partial \theta} \,
,
\end{eqnarray}
\end{widetext}
notice that in the last expression equation~(\ref{Om-exp}) has been used. 

Due to the parity of the tensor harmonics, the nonradial perturbations
can be divided in two classes, the axial (odd-parity) and polar
(even-parity) perturbations.  The polar sector can be described by
four perturbation functions: the perturbation of the relativistic enthalpy $H(r,t)$, 
the two velocity perturbations $u_1 \left( t,r\right)$, $u_2 \left( t,r\right)$, 
and the $r$ component of the displacement vector, $\xi^r(r,t)$. On the other hand,
in the Cowling approximation, axial perturbations will be described 
only by the velocity perturbation function $u_3 \left( t,r\right)$. 
Using the perturbed energy-momentum conservation equations  $\delta \left( T_{\al
\beta}^{\, \, \, \, \, \, ; \beta}\right) = 0$ one can derive a system of 
coupled differential equations for $H_{lm}$ and the $u_{i,lm}$ which together 
with equation~(\ref{xi_r_eqn}) form a complete linear system of evolution 
equations for the five unknown perturbation functions.
The four components of the perturbed energy-momentum conservation equations get the form:
\begin{widetext}
\begin{eqnarray}
\left[ t \right] && 
                 A_{lm}^{(t)} \YY + \left[ B_{1, lm}^{(t)} + B_{2, lm}^{(t)} \sin^2 \theta  \right]  \Yph  =  0 \, , 
                      \label{Ttt} \\
	       \nn \\
\left[ r \right] && 
                 A_{lm}^{(r)} \YY + \left[ B_{1, lm}^{(r)} + B_{2, lm}^{(r)} \sin^2 \theta  \right]  \Yph
               + \left[ C_{1, lm}^{(r)} + C_{2, lm}^{(r)} \sin^2 \theta  \right]  \sin \theta \, \Yth =  0 \, , \\
	       \nn \\
\left[ \theta \right] && 
                 \left[ a_{1, lm} + a_{2, lm} \cos \theta 
               + a_{3, lm} \sin ^2 \theta + a_{4, lm}^{(\theta)}  
               \cos \theta  \sin ^2 \theta \right]  \Yth 
                 \nn \\ 
              & - &  \left[ b_{1, lm} + b_{2, lm} \cos \theta 
               + b_{3, lm} \sin^2 \theta + b_{4, lm}^{(\theta)}  
               \cos \theta  \sin ^2 \theta \right] \frac{1}{\sin \theta }  \Yph = 0  \, , \\
	       \nn \\
\left[ \phi \right] &&  \left[ b_{1, lm} + b_{2, lm} \cos \theta 
               + b_{3, lm} \sin^2 \theta + b_{4, lm}^{(\phi)}  \cos \theta  \sin ^2 \theta \right]  
               \Yth \nn \\ 
              &+ &  \left[ a_{1, lm} + a_{2, lm} \cos \theta 
               + a_{3, lm} \sin^2 \theta + a_{4, lm}^{(\phi)}  
               \cos \theta  \sin ^2 \theta \right]  \frac{1}{\sin \theta }   \Yph 
               + \left[ c_{1, lm}^{(\phi)} + c_{2, lm}^{(\phi)} \sin^2 \theta \right]   
                \sin \theta \,  \YY  = 0   \, , \label{Tphph} 
\end{eqnarray}
\end{widetext}
where all the coefficients $A_{lm}$, $B_{lm}$, $C_{lm}$, $a_{i,lm}$, $b_{i,lm}$ and 
$c_{i,lm}$ are analytically written in appendix~\ref{Perturb-terms}.
Notice that in equations~(\ref{xi_r_eqn}) and~(\ref{Ttt})-(\ref{Tphph}), the
background variables $\omega$ and $\Omega$ are expanded up to the
harmonic index $l=3$. The angular dependence of the above system of 
equations is removed by performing angular integrations.  As a result, we obtain the following
system of coupled evolution equations for $H_{lm}$, $u_{1, lm}$ $u_{2,
lm}$, $u_{3, lm}$ and $\xi^r_{lm}$:
\begin{widetext}
\begin{eqnarray}
 A_{lm}^{(t)}  & + & \I m \left[ B_{1, lm}^{(t)} +  
 \mathcal{L}  _{1}^{\pm 2} B_{2, lm}^{(t)} \right]  =  0 \, , \label{Ttt-ev}\\
	       \nn \\
                 A_{lm}^{(r)}  & + & \I m \left[ B_{1, lm}^{(r)} + \mathcal{L}  _{1}^{\pm 2} B_{2, lm}^{(r)}  \right]
               + \left[ \mathcal{L}  _{1}^{\pm 1}  C_{1, lm}^{(r)} 
               +  \mathcal{L}  _{1}^{\pm 3} C_{2, lm}^{(r)} \right]   =  0 \, ,  \\
	       \nn \\
   \Lambda a _{1, lm}  & - & \I m \left[   b_{2, lm} 
                         - 2 \mathcal{L}  _{4}^{\pm 1}   b_{3, lm} 
                 + \mathcal{L}  _{2}^{\pm 2} b_{4, lm}^{(\theta)} + \mathcal{L} _{3}^{\pm 2} b_{4, lm}^{(\phi)} 
                         + c_{1, lm}^{(\phi)} +  \mathcal{L}  _{1}^{\pm 2}  
                        c_{2, lm}^{(\phi)}  \right]    \nn \\  
                       & + &\mathcal{L} _{3}^{\pm 1} a_{2, lm}  
                         +     \mathcal{L} _{4}^{\pm 2} a_{3, lm}  
                         +      \mathcal{L} _{2}^{\pm 3} a_{4, lm}^{(\theta)} 
                         +  m^2 \mathcal{L} _{4}^{\pm 1} a_{4, lm}^{(\phi)}  = 0  \, , \\
	       \nn \\
    \Lambda b _{1, lm}  & + & \I m \left[   a_{2, lm} 
                         - 2 \mathcal{L}  _{4}^{\pm 1}   a_{3, lm} 
                         + \mathcal{L} _{3}^{\pm 2} a_{4, lm}^{(\theta)} 
                         + \mathcal{L} _{2}^{\pm 2} a_{4, lm}^{(\phi)}   \right]   \nn \\
                       & + & \mathcal{L} _{3}^{\pm 1} b_{2, lm} 
                         + \mathcal{L} _{4}^{\pm 2} b_{3, lm}
                         + \mathcal{L} _{2}^{\pm 3} b_{4, lm}^{(\phi)}  
                         +  m^2 \mathcal{L} _{4}^{\pm 1} b_{4, lm}^{(\theta)} 
                         +       \mathcal{L} _{2}^{\pm 1} c_{1, lm}^{(\phi)} 
                         +       \mathcal{L} _{3}^{\pm 3} c_{2, lm}^{(\phi)}   
                      = 0.  \,  \\ 
              \nn \\ 
   \dot{\xi}_{lm} & + &\I m \left( \Om_{10} + 6 \Om_{30} \right ) \xi_{lm}
                 =  e^{2\nu-2\lambda} \left( 1 - \frac{\Gamma_1}{\Gamma} \right) \nu' u_{1,lm} 
                 +  \frac{15}{2} \Om_{30} \mathcal{L}_1^{\pm 2} \xi_{lm} \, . \label{Eq_xi}
\end{eqnarray}
\end{widetext}
where $\Lambda=l(l+1)$.
The operators $\mathcal{L}_{i}^{\pm j}$ couple perturbations
with different harmonic indeces $l$ and they are defined in
appendix~\ref{Integrals}.  The final equations can be written in a more
compact form as, 
\begin{align}
\label{polar-led}
& P_{lm} + \I m \left( P_{lm} + \tilde{A}_{l \pm 1,m} + \tilde{P}_{l \pm
2,m} \right) + A_{l \pm 1,m} + \tilde{A}_{l \pm 3,m} = 0 \, , \\ 
\label{axial-led}
& A_{lm} + \I m \left( A_{lm} + \tilde P_{l \pm 1,m} + \tilde A_{l \pm 2,m} \right)
+ P_{l \pm 1,m} + \tilde P_{l \pm 3,m} = 0 \, ,
\end{align} 
where $P_{lm}$ and $A_{lm}$ represent polar and axial
perturbation functions respectively. 
The tilded quantities denote the extra coupling terms introduced by the
differentially rotating background. It is worth noticing that apart from the $l\pm 1$ 
couplings existing in the case of uniform rotation we have extra couplings 
$l\pm 2$ and $l\pm 3$ introduced by the differential rotation. 

According to the classification of~\cite{Lockitch:2000aa} the above system
of infinite coupled equations consists of two decoupled sub-systems among which 
there is no energy exchange.  
The first one is the so called ``axial-led'' system and consists of
the polar radial $l=0$ perturbations that couple with the axial non-radial $l=1$, 
the polar non-radial $l=2$ and so on. 
The second system is the so called ``polar-led'' and consists of the polar 
dipole $l=1$ perturbations that couple with the axial non-radial $l=2$, 
the polar non-radial $l=3$ and so on. 
Since the two sub-systems are decoupled in order to excite both we have to give 
independent appropriate initial data for time evolutions. 
In our study, for both polar and axial functions with $l=2$ we assumed a 
generic gaussian pulse an an initial perturbation. 
The boundary conditions on the stellar surface are set by the vanishing of the 
Langrangian perturbation of the pressure $\Delta P =0$.

\begin{table}[!t]
\begin{center}  
\begin{ruledtabular}
\begin{tabular}{*{3}{c}}
  $l_{\rm max}$   &  F (kHz) & $H_1$ (kHz) \\
\hline 
    0   &  2.687  & 4.551 \\
    1   &  2.710  & 4.571 \\
    2   &  2.712  & 4.575 \\
    3   &  2.712  & 4.575 \\
\end{tabular}
\end{ruledtabular}
\caption{ Oscillation frequencies of the quasi-radial F and
$H_1$ modes for the B1 stelar model. With $l_{\rm max}$ we denote the maximum
harmonic index that has been used in the estimation. It is apparent that there 
is a fast convergence towards the exact value for relative small values of $l_{\rm max}$.  
\label{tab:f0} }
\end{center}
\end{table}

\section{Numerical Results} \label{Sec-Num-Res}

The perturbation equations derived in the previous sections have been studied
numerically in order to compare the accuracy of the present approach against 
published results by nonlinear evolution codes \cite{Stergioulas:2003ep}.
More specifically we study the spectral properties of axisymmetric ($m=0$)
nonradial oscillations of a slowly and differentially rotating star,
where the fluid is described by a barotropic EoS ($\Gamma_1=\Gamma$).  
An extensive study of non-axisymmetric perturbations of differentially
rotating relativistic stars for a wide range of EoS will be the subject of
a follow up paper \cite{Passamonti:nonaxy}.
In the Cowling approximation, for barotropic and axisymmetric ($m=0$)
perturbations the system of evolution
equations~(\ref{Ttt-ev})-(\ref{Eq_xi}) simplifies considerably and reduces to the following four
coupled equations:
\begin{eqnarray}
 \dot{H}_{l}  
                  & = &  \left\{ \left[  \left( \frac{2}{r} - \la ' + 2 \nu' \right) c_s^2 
                   - \nu ' \right] u_{1, \, l} + c_s^2 u_{1, \, l} '  \right\}  
                   e^{2\left( \nu - \la \right) } 
                   - c_s^2 \Lambda  \frac{e^{2\nu}}{r^2} u_{2, \, l}  \, ,  \label{axy-Heq} \\ 
\dot{u}_{1, \,l} & = &   H_{l}'  +  \left\{ \left[ 2 \left( \frac{1}{r} - \nu' \right) ( \varpi_1 + 6 \varpi_3 )
                         - \omega_1' - 6 \omega_3' \right]  \mathcal{L}  _{1}^{\pm 1}  \right. 
                      -\left. \frac{15} {2} \left[ 2 \left( \frac{1}{r} - \nu' \right) \varpi_3 - \omega_3' \right] 
                      \mathcal{L}  _{1}^{\pm 3}   \right\} u_{3, \, l}  \, , \label{axy-u1} \\ 
\dot{u}_{2, \,l}  & = &  H_{l} +  \frac{1}{\Lambda} \left\{ 2 ( \varpi_1 + 6 \varpi_3 ) \mathcal{L} _{3}^{\pm 1} 
                                  - 15 ( \Om_3 - 2 \om_3 ) 
                         \mathcal{L} _{2}^{\pm 3} \right\}  u_{3, \, l} \, , \label{axy-u2} \\ 
\dot{u}_{3, \,l}  & = &  -  \frac{1}{\Lambda} \left\{ 2 (\varpi_1 + 6 \varpi_3 ) \mathcal{L}_{3}^{\pm 1} 
                             - 30 \varpi_3 \mathcal{L}_{2}^{\pm 3} \right\} u_{2, \, l} 
                      - \frac{r}{\Lambda} \,  e^{-2\lambda} 
                               \left\{ \left(  \left( 2 - 2 r \nu' \right)  ( \varpi_1 + 6\varpi_3 )
                            + r ( \varpi_1' + 6 \varpi_3' ) \right] \mathcal{L}_{2}^{\pm 1} \right. \nn \\ 
                  & - &  \left. \frac{15}{ 2} \left[ 
                        \left( 2 - 2 r \nu' \right) \varpi_3  
                        + r \varpi_3' \right] \mathcal{L}_{3}^{\pm 3} \right\}  u_{1, \, l} \, ,  
\label{axy-u3}                     
\end{eqnarray}
where $u_{3,l}$ has been redefined as follows:
\begin{equation}
\label{u3tilde}
{u}_{3,l} = u_{3,l} - \frac{1}{\Lambda} r^2 e^{-2\nu} 
\left[ ( \varpi_1 + 6 \varpi_3 ) \mathcal{L}_2^{\pm 1} - \frac{15}{2} \varpi_3 \mathcal{L}_3^{\pm 3} \right] H_{l} \, . 
\end{equation}
Equation~(\ref{Eq_xi}) of the variable $\xi^r$ becomes trivial for barotropic pulsations, where
$\Gamma_1 = \Gamma$. In fact $\xi^r$ is no longer an independent
dynamical variable as one can argue from equations~(\ref{Bar_rel})
and~(\ref{xir-exp}).

The polar axisymmetric eigenfrequencies are symmetric 
with respect to the reversal of rotation direction, suggesting  that the rotational
corrections appear at second order in
$\Omega$~\cite{Hartle:1972ht}. Therefore, for first order slow
rotation approximation the axisymmetric modes have the same
frequencies as the nonrotating case. However, in
equations~(\ref{axy-Heq})-(\ref{axy-u3}) implicitly show up some ${\cal O}
(\Omega^2)$ coupling terms, such as the last term of
equations~(\ref{axy-u1})-(\ref{axy-u2}).  In fact in accordance with
equation~(\ref{axy-u3}), the axial velocity $u_{3,l}$ is a
dynamical variable at ${\cal O}(\Omega)$ perturbative order, while for
a nonrotating star it is stationary and its profile can be chosen on
the initial time slice. It is evident that this stationary initial
condition cannot change the mode frequencies of polar perturbations.
Here we use equations~(\ref{axy-Heq})-(\ref{axy-u3})
without discarding the $u_{3,\, l}$ terms in
equations~(\ref{axy-u1})-(\ref{axy-u2}), in order to investigate the
${\cal O}(\Omega^2)$ corrections of the axial velocity on the polar
mode frequencies. However, a complete second order slow rotation
approximation is needed for a proper analysis that will  take into
account all  coupling terms.

The system of four coupled equations~(\ref{axy-Heq})-(\ref{axy-u3})
has been evolved using second order accurate differencing methods both
in time and in space based on the two step Lax-Wendroff algorithm. As
a result, the numerical code achieved second order convergence.
As initial conditions, for the enthalpy and the three velocity
perturbations we set a Gaussian pulse, which excites several
nonradial modes. The frequencies of these nonradial modes are then
estimated by carrying out a Fast Fourier Transformation~(FFT) of the
resulting timeseries.

In general, the system of perturbation equations has an infinite
series of coupling terms, which depend on $l$. However, we got
converging numerical results by keeping a finite number of equations,
i.e. up to an $l=l_{\rm max}$. In fact, starting from a simple
configuration, where only a particular $l$ is considered, one can
gradually add up extra coupling equations until the results show
convergence.  An example is shown in table~\ref{tab:f0}, where we
tabulated the frequencies of the fundamental quasi-radial mode ($F$)
and the first overtone $H_1$ for several simulations assuming
different $l_{\rm max}$. The initial perturbed configuration is
described by spherical oscillations, i.e. $l_{\rm max} = 0$. By
gradually introducing the dipole $l_{\rm max}=1$ and quadrupole
$l_{\rm max}=2$ oscillations, the quasi-radial mode frequencies have
been improved. No further changes have been observed, when $l_{\rm
max}=3$ (or higher) perturbations has been taken into account.
Actually, in this case, convergence has been achieved already for
$l_{\rm max} =2$.  Since this behavior is shared also by the other
nonradial modes, we restricted our study up to $l_{\rm max} = 2$.

\begin{table}[!t]
\begin{center}  
\begin{ruledtabular}
\begin{tabular}{c | c | c | c | c }
  Model   &  F (kHz) & $H_1$ (kHz) &  ${}^2 f$ (kHz) &  ${}^2 p_1$ (kHz)  \\
\hline 
    B0    &  2.706 (0\%)  & 4.547 (0\%) & 1.846 (0\%)   &  4.100 (0\%)    \\
    B1    &  2.712 (2\%)  & 4.555 (2\%) & 1.895 ($<$1\%)  &   4.117 (1\%)    \\
    B3    &  2.735 (4\%)  & 4.578 (4\%) & 1.915 (1\%)   &  4.124 (2\%)    \\
    B6    &  2.797 (6\%)  & 4.624 (4\%) & 1.944 (1\%)   &  4.134 (7\%)    \\
    B9    &  2.885 (8\%)  & 4.686 (6\%) & 1.974 (8\%)  &   4.147 (14\%) \\
\end{tabular}
\end{ruledtabular}
\caption{\label{tab:Modes} Frequencies of the first two quasi-radial
($F$ and $H_1$) and nonradial ($^2f$ and $^2p_1$) modes, for differentially 
rotating B-models with $A=12$ and $l_{\rm max} = 2$.  
The relative difference between the perturbative and nonlinear
results~\cite{Stergioulas:2003ep} is shown in parenthesis}.
\end{center}
\end{table}

The accuracy of our description has been tested against published
results from a nonlinear code ~\cite{Stergioulas:2003ep}~(SAF from now
on) for a  sequence of stellar models.  The frequencies of the
first two quasi-radial ($F$ and $H_1$) and two $l=2$ nonradial modes
($^2f$ and $^2p_1$ ) for some of the B-models are listed in
table~\ref{tab:Modes}.  Although these stellar models describe very
fast rotating stars (see table \ref{tab:Models}), the perturbative
results show a good agreement (within the limits of the slow rotation approximation)
with those obtained with nonlinear
simulations in SAF.  In fact, a difference of the order of $7$\%
appears only after the B6 model.  Errors of this order are expected
for very fast rotating stars since our results, which are based on the
slow rotation approximation (first order in $\Omega$), cannot describe
the flattening of the star due to the centrifugal force.

An important aspect to emphasize is that our results do not show any
splitting of the fundamental quasi-radial mode, as observed in SAF.
This result supports the argument in~\cite{Dimmelmeier:2005zk},  
that this effect (splitting) could be an artifact of the Cowling approximation 
in the  non-linear regime. Moreover, we don't observe any significant change
in the frequency of the $^2p_1$-mode in constrast to the results in SAF where 
an overall change of the order of 30\% has been observed.

The degree of differential rotation affects the mode frequencies
due to the presence of extra coupling terms in the perturbation
equations~(\ref{Ttt-ev})-(\ref{Eq_xi}). 
This effect can then be studied for several B-stellar models by
varying the parameter $A$, from a highly differentially rotating
configuration with $A=12$~km to a uniformly rotating star with a very large $A$
e.g. $A=500$~km. 
As shown in figure~\ref{tab:f2-mode}, the frequency of the nonradial
 fundamental ${}^2 f$ mode increases as the parameter $A$ increases
 for the whole sequence of B-models.  In particular, the ${}^2 f$
 frequency already converges to its value for uniform rotation when $A
 \approx 100$~km. In this case, the angular velocity at the equator
 $\Omega_e$ is typically about 2.5\% smaller than $\Omega_c$ on the
 rotational axis.
It is worth noticing that the effect of $A$ on the mode frequencies is
considerably larger for fast rotating stellar model.  As shown in
table~\ref{tab:f2-mode}, the same behavior has been observed also for
the other quasi-radial and nonradial modes. Among these modes, the
$F$-mode is the most sensitive to the differential rotation, as it can
change up to 25\% for the fastest rotating model B9. Instead, the
${}^2 f$ and ${}^2 p_1$ modes can change up to 10 percent and 3
percent respectively. 

\begin{figure}[!t]
\begin{center}
\includegraphics[width=95mm,height=85mm]{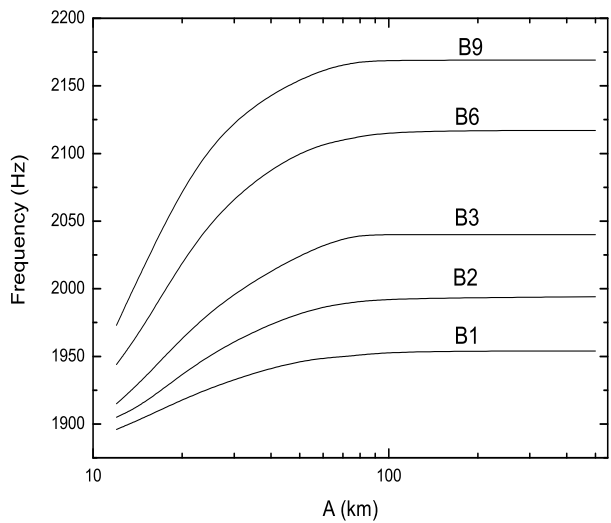}
\caption{ The frequency of the ${}^2 f$ mode as function of the parameter $A$
is plotted for various B-models. For $A=500~{\rm km}$ the star is practically
rotates uniformly. The shifting of the mode frequency between the maximum and 
minimum values of A varies from 3\% to 10\% depending on the model.
\label{fig:Freqs_B2}}
\end{center}
\end{figure}

\begin{table}[!t]
\begin{center}  
\begin{ruledtabular}
\begin{tabular}{c |  c |  c   c   c c c  }
   Modes          &  A\,[km]  &   B0    & B1     & B3      & B6        & B9         \\
\hline 
\hline
                  & 12       & 2.706   & 2.702  & 2.735   & 2.797   &   2.885     \\
    F             & 50       & 2.706   & 2.738  & 2.849   & 3.058   &   3.352     \\
                  & 75       & 2.706   & 2.764  & 2.954   & 3.294   &   3.747     \\
                  & 100      & 2.706   & 2.771  & 2.960   & 3.313   &   3.777     \\
\hline \hline  
                  & 12       & 1.846   & 1.895  & 1.915   & 1.944   &  1.974      \\
${}^2 f$          & 50       & 1.846   & 1.928  & 1.987   & 2.058   &  2.117      \\
                  & 75       & 1.846   & 1.947  & 2.038   & 2.111   &  2.169      \\
                  & 100      & 1.846   & 1.954  & 2.040   & 2.118   &  2.170      \\
\hline \hline 
                  & 12       & 4.100   & 4.117  & 4.124   & 4.134   &  4.147      \\
${}^2 p_1$        & 50       & 4.100   & 4.124  & 4.143   & 4.163   &  4.196      \\
                  & 75       & 4.100   & 4.130  & 4.155   & 4.175   &  4.209      \\
                 & 100       & 4.100   & 4.134  & 4.157   & 4.176   &  4.210      \\
\end{tabular}
\end{ruledtabular}
\caption{\label{tab:f2-mode} Frequencies, in kHz, of the quasi-radial
F mode, and ${}^2 f$ and ${}^2 p_1$ nonradial modes for B-stellar
models with different rotational parameter $A$.}
\end{center}
\end{table}

\section{Conclusions and Discussion\label{conclusions}}

In this article we derived the general equations describing the perturbations of
slowly and differentially rotating neutron stars in the Cowling approximation.
The equations have been derived for spherical stars (slow rotation) on which 
the differential rotation has been described via the j-constant law. 

As a first step and test, results have been derived for axisymmetric,
barotropic oscillations. In the system of perturbation
equations some terms of second order ${\cal O} (\Omega^2)$ have been maintained
and their effects on the axisymmetric eigenfrequencies has been studied. The
comparison of the derived results with those produced by the full
non-linear code \cite{Stergioulas:2003ep} suggests that this
perturbative approach can provide quite accurate estimates of the
oscillation frequencies within the limits of the slow rotation
approximation.  A point to be stressed is that since we keep only
terms first order in rotation ${\cal O}(\Omega)$ the background
stellar model is still spherical, and the compactness of the star
${M}/{R}$ remains the same along the sequence of stellar models.  This
seems to be one of the most important reasons for the deviations
observed between our results and those of SAF for the faster rotating
members of the B-model sequence. It is expected that the inclusion of
second order terms in rotation, which will take into account the
flattening of the star, will improve the agreement between the
perturbative results with the non-linear ones for the very fast
rotating models.

An advantage of this perturbative approach is the numerical simplicity and the
ability for an easier physical interpretation of the derived results. 
In a follow up paper we study the non-axisymmetric oscillations of differentially
rotating relativistic stars for various EoS with special emphasis to the hot EoS.

\[ \]
{\bf Acknowledgments:} We are grateful to Nikolaos Stergioulas,
Hajime Sotani and Miltos Vavoulidis for helpful discussions.  A.S. is
supported by the Pythagoras II grand of the Greek Ministry of Research
and Development.  A.P. is supported by a ``Virgo Ego Scientific Forum'' 
(VESF) and by the EU program ILIAS.

\appendix

\section{Harmonic components of the j-constant rotation law \label{Rot-law-Int}}

In this Appendix we present the explicit expression for the terms
$\mathcal{I}_{l}^{N}$ and $\mathcal{I}_{l}^{l'}$ described in
equations~(\ref{In-def}) and~(\ref{Jlm-def}). Since the series
expansions (\ref{om-exp})-(\ref{Om-exp}) for $\omega$ and $\Omega$
have been truncated for $l>3$, and that terms with even $l$ vanish due
to the rotation law parity, we show the expressions for harmonic
indices $l=1$ and $l=3$.  The two nearly ``Newtonian terms''
$\mathcal{I}_{l}^{N}$ are:
\begin{widetext}
\begin{eqnarray}
\mathcal{I}_{1}^{N} & = & \frac{3}{2} \frac{A^2}{\chi^2} \, \Omega_{c} \, 
                           \left[1-\frac{A^2}{\chi \sqrt{A^2 + \chi^2} }\
                           \ln\left( \frac{\sqrt{\sqrt{A^2 + \chi^2}+\chi}}
                                          {\sqrt{\sqrt{A^2 + \chi^2}-\chi}}\right)
                           \right] \,, \label{bet01_j_law_l1} \\
\mathcal{I}_{3}^{N} & = & \frac{7}{12}  \frac{A^2}{\chi^2} \, \Omega_{c} \, 
                          \left[  1 + 7.5 \frac{A^2}{\chi^2}
                           - \frac{6 A^2}{\chi\sqrt{A^2 + \chi^2}}
                              \left(1+\frac{5}{4}\frac{A^2}{\chi^2}\right)
                           \ln\left(\frac{\sqrt{\sqrt{A^2 + \chi^2}+\chi}}
                                         {\sqrt{\sqrt{A^2 + \chi^2} - \chi}}
                           \right)  \right] \, ,  \label{bet03_j_law_l1}  
\end{eqnarray}
\end{widetext}
The four terms that contribute in the estimation of the ``relativistic corrections''
$\mathcal{I}_{l}^{R}$ are:
\begin{widetext}
\begin{eqnarray}
\mathcal{J}_{1}^{1} & = &  \frac{4}{3} \frac{\pi}{\chi^4} \left[ 2 \chi^2 - 3 A^2 
                          + \frac{3 A^4}{\chi \sqrt{A^2 + \chi^2}}
                          \ln\left(\frac{\sqrt{\sqrt{A^2 + \chi^2}+\chi}}
                                        {\sqrt{\sqrt{A^2 + \chi^2}-\chi}}\right)
                          \right] \,,  \label{J1010-def} \\
\mathcal{J}_{1}^{3} & = & \mathcal{J}_{3}^{1} =  - 2 \pi  \frac{ A^2}{\chi^6} 
                                  \left[ 2 \chi^2 + 15 A^2 - 
                                  \frac{3 A^2 \left( 4 \chi^2 + 5 A^2 \right)}
                                       { \chi \sqrt{A^2 + \chi^2}}
                             \ln\left(\frac{\sqrt{\sqrt{A^2 + \chi^2}+\chi}}
                                           {\sqrt{\sqrt{A^2 + \chi^2}-\chi}}\right)
                            \right] \,,  \label{J1030-def} \\
\mathcal{J}_{3}^{3} & = &  \frac{3}{7}  \frac{\pi}{\chi^8} 
                           \left[ - 525 A^6 - 490 A^4 \chi^2  - 56 A^2 \chi^4 + 16 \chi^6  
                          + \frac{A^4}{ \chi \sqrt{A^2 + \chi^2}} 
                            \left( 525 A^4 + 840 A^2 \chi^2 + 336 \chi^4 \right) 
                            \right. {} \nn \\                           
                       && \times \left. \ln\left(\frac{\sqrt{\sqrt{A^2 + \chi^2}+\chi}}
                                                {\sqrt{\sqrt{A^2 + \chi^2}-\chi}}\right)
                          \right] \,,  \label{J3030-def} 
\end{eqnarray}
\end{widetext}
where we have set $\chi = r e^{-\nu}$.

\section{Coefficients of the perturbed conservation equations\label{Perturb-terms}}
In this Appendix we present the coefficients of the perturbation
equations~(\ref{Ttt})-(\ref{Tphph}),
\begin{eqnarray}
 A_{lm}^{(t)}    & = &  - \dot{H}_{lm} - c_s^2 l \left( l + 1 \right)  \frac{e^{2\nu}}{r^2} u_{2, \, lm} 
                   + \left\{ \left[  \left( \frac{2}{r} - \la ' + 2 \nu' \right) c_s^2 
                   - \nu ' \right] u_{1, \, lm} + c_s^2 u_{1, \, lm} '  \right\}  e^{2\left( \nu - \la \right) }\, , \\ 
     \nn \\
 B_{1, lm}^{(t)} & = &  \left[ - \Om_{1} - 6 \Om_{3} + c_s^2 \left( \varpi_{1} + 6  \varpi_{3} \right) \right] H_{lm} \, ,  \\
     \nn \\
 B_{2, lm}^{(t)} & = &  - \frac{15}{2} \left(  c_s^2 \varpi _{3} - \Om_{3} \right)  H_{lm} \, ,  \\
 A_{lm}^{(r)}    & = &  - \dot{u}_{1, \,lm} + H_{lm}' + \nu' c_s^{-2} \left[ \xi_{lm}  
                      + \left( 1 - \frac{\Gamma_1}{\Gamma} \right) H_{lm} \right]  \, , \\
     \nn \\
 B_{1, lm}^{(r)} & = &   \left\{ \left[ 2 \left( \frac{1}{r} - \nu' \right) \varpi_{1} - \om _{1}' \right]   
                        +  6 \left[ 2 \left( \frac{1}{r} - \nu' \right) \varpi_{3}
                        -  \om _{3}' \right] \right\} u_{2, \, lm} \nn \\ 
                 &  - & \left( \Om_{1} + 6 \Om _{3} \right) u_{1, \, lm}    \, , \\
     \nn \\
 B_{2, lm}^{(r)} & = &   \frac{15}{2} \Om _{3} u_{1, \, lm}  - \frac{15}{2} \left[ 2 \varpi_{3} 
                          \left( \frac{1}{r} - \nu' \right) -  \om _{3}' \right] u_{2, \, lm} \, ,\\
 C_{1, lm}^{(r)} & = &   \left\{ \left[ 2 \left( \frac{1}{r} - \nu' \right) \varpi_{1}
                         - \om _{1}' \right]  
                        + 6 \left[ 2 \left( \frac{1}{r} - \nu' \right) \varpi_{3}
                        - \om _{3}' \right] \right\} u_{3, \, lm} \,\\ \nn \\
 C_{2, lm}^{(r)} & = &     - \frac{15}{2}  \left[ 2 \varpi_{3} \left( \frac{1}{r} - \nu' \right) 
                         -  \om _{3}' \right] u_{3, \, lm} \, , 
\end{eqnarray}
\begin{eqnarray}
 a_{1, lm}          & = &  - \dot{u}_{2, \,lm} + H_{lm} - \I m \left( \Om _{1} + 6 \Om _{3} \right) u_{2,\,lm}  \, , \\ \nn \\
 a_{2, lm}          & = &   2 \left(  \varpi _{1} + 6  \varpi_{3} \right) u_{3, \, lm} \, , 
      \\ \nn \\
 a_{3, lm}          & = &    \frac{15}{2} \, \I m \, \Om _{3} \, u_{2, \, lm}   \, , 
       \\  \nn \\ 
 a_{4, lm}^{(\theta)} & = &    - 15 \left( \Om _{3} - 2 \om_{3} \right) \, u_{3, \, lm} \, , 
      \\ \nn \\
 a_{4, lm}^{(\phi)}   & = &    - 30 \varpi _{3} \,  u_{3, \, lm} \, , 
      \\ \nn \\
 b_{1, lm}          & = &  - \dot{u}_{3, \,lm}  
                      - \I m \left( \Om _{1} + 6 \Om _{3} \right) u_{3, \, lm} 
        \\  \nn \\ 
 b_{2, lm}          & = &  - 2 \left[  \varpi _{1} + 6 \varpi _{3}  \right] u_{2, \, lm} \, , 
      \\ \nn \\
 b_{3, lm}          & = &   \frac{15}{2} \, \I m \, \Om _{3} \, u_{3, \, lm}  \, , 
       \\  \nn \\ 
 b_{4, lm}^{(\theta)} & = &  15 \left( \Om _{3} - 2 \om_{3} \right) \, u_{2, \, lm} \, , 
       \\  \nn \\ 
 b_{4, lm}^{(\phi)}   & = &  30 \varpi_{3} \, u_{2, \, lm} \, , 
       \\  \nn \\ 
 c_{1, lm}^{(\phi)} & = &   c_s^2 \left( \varpi_{1} + 6 \, \varpi_{3} \right) 
                          \left(  r^2   e^{-2 \lambda} \, u_{1, \, lm}'   -\Lambda u_{2, \, lm} \right)   \nn \\   
                  & - &  r e^{-2 \lambda} \left\{ \left[ 2 - r \nu ' - c_s^2 \left( 2 + r \left( 2 \nu ' - \lambda ' \right) 
                      \right)  \right] \left( \varpi_{1} + 6 \, \varpi_{3} \right)
                         +  r \left( \varpi'_{1} + 6 \, \varpi'_{3} \right) \right\} u_{1, \, lm} \nn \\
                  & = & r^2 e^{-2\nu} \left( \varpi_1 + 6\varpi_3 \right) \dot{H}_{lm}
                        - re^{-2\lambda} \left[ ( 2 - 2 r \nu' ) ( \varpi_1 + 6 \varpi_3 )
                               + r( \varpi' _1 + 6 \varpi ' _3)\right] u_{1,lm},  \\
 c_{2, lm}^{(\phi)} & = &   \frac{15}{2}  c_s^2 \, \varpi_{3} \left( \Lambda \, u_{2, \, lm}  
                        - r^2  e^{-2 \lambda} u_{1, \, lm}' \right)                  \nn \\ 
                  & + &   \frac{15}{2} r \, e^{-2 \lambda} \left\{ \left[ 2 - r \, \nu ' - c_s^2 \left( 
                              2 + r \left( 2 \nu' - \lambda' \right) \right) \right]  
                             \varpi _{3} + r \varpi '_{3} \right\} u_{1, \, lm}    \nn \\
                  & = & -\frac{15}{2} r^2 e^{-2\nu} \varpi_3 \dot{H}_{lm} 
                        + \frac{15}{2} r e^{-2\lambda} 
                           \left[  ( 2 - 2 r\nu' )  \varpi_3 + r \varpi' _3  \right] u_{1,lm},
\end{eqnarray}
where we have used the definitions,
\begin{eqnarray}
\varpi_{1}(r) &\equiv& \Om_{1}(r) - \om_{1}(r) \nn \, ,\\
\varpi_{3}(r) &\equiv& \Om_{3}(r) - \om_{3}(r) \nn \, .
\end{eqnarray}

\section{Angular Integration \label{Integrals}}

Here, we define the linear operators $\mathcal{L}_{i}^{\pm j}$ with
$i,j \in \mathbb{N}$, which come out from the angular integration of
equations~(\ref{Ttt})-(\ref{Tphph}). This procedure relies on the
orthogonality properties of the spherical harmonics, on the following angular
integrals:
\begin{widetext}
\begin{eqnarray}
 \int _{S^2} d \tilde{\Om} \left( \frac{\partial Y_{l'm'}^{\ast} }{\partial \theta} 
                                 \frac{\partial Y_{lm} }{\partial \theta}  
                        + \frac{1}{\sin ^2 \theta} \frac{\partial Y_{l'm'}^{\ast} }{\partial \phi} 
                           \frac{\partial Y_{lm} }{\partial \phi} \right) 
              &  = &  l \left( l +1 \right) \delta_{ll'} \delta_{mm'}  \, , 
           \\ \nn \\ 
 \int _{S^2} d \tilde{\Om} \left(  Y_{l'm'}^{\ast} \frac{\partial Y_{lm} }{\partial \theta}  
                              + \frac{\partial Y_{l'm'}^{\ast} }{\partial \theta} Y_{lm}  
                           \right) \frac{\cos \theta}{\sin \theta} 
              &  = &  \delta_{ll'} \delta_{mm'}  \, ,  
\end{eqnarray}
\end{widetext}
and the two relations:
\begin{eqnarray}
\cos \theta \, \YY    &=&   Q_{l+1 m} Y_{l+1 m} + Q_{l m} Y_{l-1 m}  \, ,\\ 
\sin \theta \, \partial_\theta\YY   &=&   Q_{l+1 m} l Y_{l+1 m} - Q_{l m} 
\left( l +1 \right) Y_{l-1 m} \, , 
\end{eqnarray}
where $Q_{lm}$ is defined as follows:
\begin{equation}
Q_{lm} = \sqrt{\frac{\left( l-m \right) \left( l+m \right)}{\left(2l-1\right) \left(2l+1\right)}} \, .
\end{equation}
The operators $\mathcal{L}_{j}^{\pm 1}$ which introduce couplings of a perturbation
term/expression with harmonic index $l$ with the term/expression having 
harmonic indices $l\pm 1$ are given by the following relations:
\begin{widetext}
\begin{eqnarray}
\mathcal{L}  _{1}^{\pm 1}  \mathcal{A} & \equiv & \sum_{l'm'} \mathcal{A}_{l'm'} 
                   \int_{S^2} d \tilde{\Om} \, Y_{lm}^{\ast}  \sin   \theta \, \partial_{\theta} Y_{l'm'} 
                     =  \left( l - 1 \right) Q_{l m} \mathcal{A}_{l-1 m} 
                          - \left( l + 2 \right) Q_{l+1 m} \mathcal{A}_{l+1 m} \, , \\ \nn \\
\mathcal{L}  _{2}^{\pm 1}  \mathcal{A} & \equiv & \sum_{l'm'} \mathcal{A}_{l'm'} 
                   \int_{S^2} d \tilde{\Om} \, \partial_{\theta} Y_{lm}^{\ast}  \sin\theta \, Y_{l'm'} 
                   = - \left( l+1 \right) Q_{l m} \mathcal{A}_{l-1 m} + l Q_{l+1 m} \mathcal{A} _{l+1 m} \, , \\ \nn \\ 
\mathcal{L}  _{3}^{\pm 1}  \mathcal{A} & \equiv & \sum_{l'm'} \mathcal{A}_{l'm'} 
                   \int_{S^2} d \tilde{\Om} \,  \left(  \partial_{\theta}  Y_{lm}^{\ast}  \, \partial_{\theta} Y_{l'm'}  
                        + \frac{1}{\sin ^2 \theta} \partial_{\phi} Y_{lm}^{\ast} \, 
                       \partial_{\phi} Y_{l'm'} \right) \cos \theta \nn \\
                   &=&  \left( l -1 \right) \left( l + 1 \right) Q_{lm} \mathcal{A}_{l-1 m} 
                    +  l \left( l + 2 \right) Q_{l+1 m} \mathcal{A}_{l+1 m}   \, , \\ \nn \\ 
\mathcal{L}  _{4}^{\pm 1}  \mathcal{A} & \equiv & \sum_{l'm'} \mathcal{A}_{l'm'} 
                   \int_{S^2} d \tilde{\Om} \, Y_{lm}^{\ast}  \cos \theta \, Y_{l'm'} 
                   = Q_{lm} \mathcal{A}_{l-1 m}  + Q_{l+1 m} \mathcal{A}_{l+1 m}   \, , \\ \nn \\ 
\mathcal{L}  _{5}^{\pm 1}  \mathcal{A} & \equiv & \sum_{l'm'} \mathcal{A}_{l'm'}  
                   \int_{S^2} d \tilde{\Om} \, \left( 
                        \partial_{\theta}  Y_{lm}^{\ast}  \, \partial_{\phi}  Y_{l'm'}  -  
                        \partial_{\phi}  Y_{lm}^{\ast}  \, \partial_{\theta}  Y_{l'm'}  \,  
                       \right) \sin \theta   
                   = \I m \left( \mathcal{L}  _{1}^{\pm 1} + \mathcal{L}  _{2}^{\pm 1} \right)
                  \, . \\ \nn 
\end{eqnarray}
\end{widetext}
Notice that the operator $\mathcal{L} _{4}^{\pm 1}$ can be defined in
terms of the $\mathcal{L} _{1}^{\pm 1}$ and $\mathcal{L} _{2}^{\pm 1}$
operators as follows:
\begin{equation}
\mathcal{L}  _{4}^{\pm 1} = -\frac{1}{2} ( \mathcal{L}  _{1}^{\pm 1} + \mathcal{L}  _{2}^{\pm 1} ) \nonumber 
\end{equation}
In a similar way the operators $\mathcal{L}_{j}^{\pm 2}$ that introduce couplings of a 
perturbation expression with harmonic index $l$ with term/expression with 
harmonic indices $l\pm 2$ are: 
\begin{widetext}
\begin{eqnarray}                                                  
\mathcal{L}  _{1}^{\pm 2}  \mathcal{A} & \equiv & \sum_{l'm'} \mathcal{A}_{l'm'} 
                   \int_{S^2} d \tilde{\Om} \, Y_{lm}^{\ast}  \sin^2 \theta \, Y_{l'm'} 
                     =   - Q_{l-1 m} Q_{l m} \mathcal{A}_{l-2 m}  + \left( 1 - Q_{lm}^{2} - Q_{l+1 m}^{2}\right) 
                      \mathcal{A}_{lm}  \nn \\ && -  Q_{l m} Q_{l+1 m} \mathcal{A}_{l+2 m}       \, , \\ \nn \\
\mathcal{L}  _{2}^{\pm 2}  \mathcal{A} & \equiv & \sum_{l'm'} \mathcal{A}_{l'm'} 
                   \int_{S^2} d \tilde{\Om} \, \partial_{\theta} Y_{lm}^{\ast}  \sin \theta \cos \theta \, Y_{l'm'} 
                   = - \left( l+1 \right) Q_{l-1 m} Q_{lm} \mathcal{A}_{l-2 m} 
                     + \left[ l Q_{l+1 m}^{2} - \left( l+1 \right) Q_{lm}^{2} \right] \mathcal{A}_{lm} \nn \\ 
                   && + l Q_{l+1 m} Q_{l+2 m} \mathcal{A}_{l+2 m}   \, , \\ \nn \\ 
\mathcal{L}  _{3}^{\pm 2}  \mathcal{A} & \equiv & \sum_{l'm'} \mathcal{A}_{l'm'} 
                   \int_{S^2} d \tilde{\Om} \, Y_{lm}^{\ast}  \sin \theta \cos \theta \, \partial_{\theta} Y_{l'm'} 
                   =    \left( l-2 \right) Q_{l-1 m} Q_{lm} \mathcal{A}_{l-2 m}  
                      + \left[ l Q_{l+1 m}^{2} - \left( l+1 \right) Q_{lm}^{2} \right] \mathcal{A}_{lm} \nn \\
                   && - \left( l +3 \right)  Q_{l+1 m} Q_{l+2 m} \mathcal{A}_{l+2 m}    \, ,  \\ \nn \\ 
\mathcal{L}  _{4}^{\pm 2}  \mathcal{A} & \equiv & \sum_{l'm'} \mathcal{A}_{l'm'} 
                   \int_{S^2} d \tilde{\Om} \,  \left(  \partial_{\theta}  Y_{lm}^{\ast}  \, \partial_{\theta} Y_{l'm'}  
                        + \frac{1}{\sin ^2 \theta} \partial_{\phi} Y_{lm}^{\ast} \, 
                        \partial_{\phi} Y_{l'm'} \right) \sin^2 \theta
                   = - \left(l-2\right) \left(l+1\right) Q_{l-1 m}Q_{lm} \mathcal{A}_{l-2 m} \nn \\
                   && + \left[ l \left(l+1\right) - \left( l+1\right) \left( l-2\right) Q_{lm}^{2} 
                      - l \left( l+3 \right) Q_{l+1 m}^{2} \right] \mathcal{A}_{lm} 
                      - l \left( l+3 \right) Q_{l+1 m} Q_{l+2 m} \mathcal{A} _{l+2 m} \, . \\ \nn 
\end{eqnarray}
\end{widetext}
Finally, the operators $\mathcal{L}_{j}^{\pm 3}$ that introduce couplings of a 
perturbation expression with harmonic index $l$ with terms/expressions with 
harmonic indices $l\pm3$ are: 
\begin{widetext}
\begin{eqnarray}                                                  
\mathcal{L}  _{1}^{\pm 3}  \mathcal{A} & \equiv & \sum_{l'm'} \mathcal{A}_{l'm'} 
                   \int_{S^2} d \tilde{\Om} \, Y_{lm}^{\ast}  \sin^3 \theta \,  \partial_{\theta}  Y_{l'm'} 
                     =  - \left(l - 3 \right) Q_{l-2 m}  Q_{l-1 m} Q_{l m} \, \mathcal{A}_{l-3 m} 
                       + \left(l +4 \right) Q_{l+1 m}  Q_{l+2 m} Q_{l+3 m} \, \mathcal{A}_{l+3 m}\nn \\
                     && + Q_{lm} \left[ l  Q_{l-1 m}^{2} + \left(l-1 \right) \left( 1 -  Q_{l m}^{2} -  Q_{l+1 m}^{2} 
                        \right) \right]   \mathcal{A}_{l-1 m} 
                       \nn  \\ && 
		       - Q_{l+1 m}  \left[ \left( l + 1 \right)  Q_{l+2 m}^{2}  + \left( l+2\right) 
                               \left(1 -  Q_{l m}^{2} - Q_{l+1 m}^{2} \right) \right]  \mathcal{A}_{l+1 m} \, , 
                      \\ \nn \\ 
\mathcal{L}  _{2}^{\pm 3}  \mathcal{A} & \equiv & \sum_{l'm'} \mathcal{A}_{l'm'} 
                   \int_{S^2} d \tilde{\Om} \, \partial_{\theta} Y_{lm}^{\ast}  \cos \theta \sin^2 \theta \, 
                                                                                             \partial_{\theta} Y_{l'm'} 
                   = - \left( l - 3\right) \left( l+1\right) Q_{l-2 m} Q_{l-1 m} Q_{l m} \mathcal{A}_{l-3 m} \nn \\  
                   &&  + Q_{lm} \left[  l \left( l+1 \right) Q^2_{l-1 m} - \left( l-1\right) \left( l+1\right) Q_{l m}^{2} 
                     +  l \left( l-1 \right) Q^2_{l+1 m} \right] \mathcal{A}_{l-1 m} \nn \\
		   &&   + Q_{l+1 m} \left[  \left( l+1 \right) \left( l+2 \right) Q^2_{lm} 
                     - l \left( l+2\right) Q_{l+1 m}^{2} +  l \left( l+1 \right) Q^2_{l+2 m} \right] 
                       \mathcal{A}_{l+1 m} \nn \\ 
                   && - l \left( l +4 \right) Q_{l+1 m} Q_{l+2 m} Q_{l+3 m} \mathcal{A}_{l+3 m}  \, , \\ \nn \\  
\mathcal{L}  _{3}^{\pm 3}  \mathcal{A} & \equiv & \sum_{l'm'} \mathcal{A}_{l'm'} 
                   \int_{S^2} d \tilde{\Om} \, \partial_{\theta} Y_{lm}^{\ast}  \sin^3 \theta \, Y_{l'm'} 
                  =     \left( l+1\right) Q_{l-2 m} Q_{l-1 m} Q_{l m} \mathcal{A}_{l-3 m} \nn \\  
                  &&  - Q_{lm} \left[  \left( l+1 \right) + l Q^2_{l+1 m} - \left( l+1\right) 
                        \left( Q_{l-1 m}^{2} + Q^2_{l  m} \right) \right] \mathcal{A}_{l-1 m} \nn \\ 
		   &&  + Q_{l+1 m} \left[ l +  \left( l+1 \right) Q^2_{lm} 
                       - l \left( Q_{l+1 m}^{2} + Q^2_{l+2 m} \right) \right] \mathcal{A}_{l+1 m} \nn \\  
                   && - l Q_{l+1 m} Q_{l+2 m} Q_{l+3 m} \mathcal{A}_{l+3 m}  \, .
\end{eqnarray}
\end{widetext}

\section{Rotation law in isotropic coordinates \label{sec:rot-law-isotr}}
In this Appendix, we describe the j-constant rotation
law in isotropic coordinates, and the appropriate transformation
in Schwarzshild coordinates.  
In isotropic coordinates the axially symmetric spacetime for a slowly
rotating body can be described by the following metric:
\begin{equation}
ds^2 = - e^{2\alpha} dt^2  +  e^{2\beta} \left[ d \hat{r}^2 + \hat{r} ^2 
                    \left(  d \theta ^2 
                    + \sin^2 \theta d \phi ^2 \right) \right]   
   -  2 \, e^{2\beta} \, \omega  \, \hat{r}^2 \sin ^2 \theta \, dt \, d \phi 
 \, ,  \label{ds-equil_iso}
\end{equation}
where $\alpha$ and $\beta$ are functions of the isotropic coordinate
$\hat{r}$. Equation~(\ref{ds-equil_iso}) has been derived by the slow
rotating metric~(\ref{ds-equil}) in Schwarschild coordinate  via
the following relations:
\begin{eqnarray}
\frac{d \hat{r}} {d r} & = & \frac{\hat{r}}{r} \left( 1 - \frac{2 M}{r} \right) ^{-\frac{1}{2}} \, , \\
 e^{\beta} & = & \frac{r}{\hat{r}}  \label{beta_S} \, ,   
\end{eqnarray}
where $M=M(r)$ is the mass function in Schwarzshild coordinates.  At
the stellar surface the metric functions must smoothly join the
Schwarzshild solution, this leads to the following matching conditions  
in isotropic coordinates 
\begin{eqnarray}
e^{\alpha} & = &  \left( 1-\frac{M}{2\hat{r}} \right) \left(1+\frac{M}{2\hat{r}}\right)^{-1} \, , 
\label{alp_S_iso}\\
e^{\beta} & = &   \left(1+\frac{M}{2\hat{r}}\right)^{2} \, . 
\label{beta_S_iso}
\end{eqnarray}
The stellar radius $\hat{R}$, in isotropic coordinates, is given in
terms of the stellar mass $M$ and the radius $R$ in Schwarzshild
coordinates by the following relation:
\begin{equation}
\hat{R} = \frac{1}{2} \left( R - M + \sqrt{ R^2 - 2 M R } \right) \,  . \label{R-iso}
\end{equation}
Finally, the relativistic j-constant rotation law can be determined via the relation:
\begin{equation}
u^t u_{\phi} = \hat{A}^2 \left( \Omega_c - \Omega \right) \, ,
\end{equation}
which leads to the following expression of the angular velocity
profile:
\begin{equation}
\Omega(\hat r,\theta) = \frac{\hat A^2 \, \Omega^{}_{c} + e^{2 \left(\beta-\alpha \right) } \, 
\hat r^2 \, \sin^2\theta \,
\omega(\hat r,\theta)}{\hat A^2 + e^{2 \left(\beta-\alpha \right)} \, \hat  r^2 \, \sin^2\theta} \, ,  
\label{j-cons-iso}
\end{equation}
where $\hat{A}$ is a parameter that describes the
degree of differential rotation in isotropic coordinates.
By defining a new function
\begin{equation}
\hat x = e^{\beta-\alpha } \, \frac{ \hat{r} }{\hat A}  \sin\theta  \, , 
\end{equation}
we can rewrite equation~(\ref{j-cons-iso}) in a more compact form i.e.
\begin{equation}
\frac{ \Omega(\hat r,\theta) }{\Omega_c }= 
           \frac{ 1  + \hat x ^2 \omega(\hat r,\theta) / \Omega_c}{1 + \hat x^2 } \, . 
         \label{j-cons-iso-red}
\end{equation}
It is worth noticing that equation~(\ref{j-cons}) can be written in
the same form of equation~(\ref{j-cons-iso-red}) if the function $\hat x$ is 
replaced by 
\begin{equation}
x = e^{-\nu } \, \frac{ r }{A} \sin\theta \, . \label{x_sch}
\end{equation} 
Equations~(\ref{j-cons-iso}) and~(\ref{j-cons}) are obviously related
by the coordinate transformation between isotropic and Schwarzshild
coordinates.  
A typical choice for $\hat{A}$  is to set its value equal
to the stellar radius $\hat R$, which, as we show later, corresponds 
to $\Omega_e \sim \Omega_c / 3 $. This relation
between the angular velocity at the equator and the rotation axis
seems to be in good agreement with the rotation patters estimated in the remnants
of hypermassive binary mergers~\cite{Shibata:1999wm}. 
The matching conditions~(\ref{alp_S_iso})
and~(\ref{beta_S_iso}) lead to the following expression for $\hat x_{e}$ 
at the equator:
\begin{eqnarray}
\hat x_{e} & = & \left(1+\frac{M}{2\hat{R}}\right)^{3} \left(
              1-\frac{M}{2\hat{R}} \right)^{-1}
              \frac{\hat{R}}{\hat{A}} \, . \label{hat_x_e} 
\end{eqnarray}
A coordinate transformation leads to the following expression for ${x_e}$
in Schwarzshild coordinates:
\begin{equation}
x_{e} =  \left( 1- \frac{2 M}{R} \right)^{-1/2} \frac{R}{\hat{A}} \, . \label{x_e} \\
\end{equation}
From equations~(\ref{j-cons-iso-red})-(\ref{x_e}), it is worth
noticing that the ratio $\Omega_e / \Omega_c$ depends practically on the
compactness of the star.  In order to set up an equilibrium
configuration in Schwarzshild coordinates, which has the same
value of $\Om_e/\Om_c$ as the one used in non-linear calculations
in isotropic coordinates, 
the natural choice is $A=\hat{A}$ as it comes out from the 
equations~(\ref{x_e}) and~(\ref{x_sch}).  
For example, for a polytropic stellar
model with mass $M=1.4~M_{\odot}$ and radius $R=14.151~{\rm km}$,
equation~(\ref{R-iso}) gives $\hat{R} =12~{\rm km}$ and leads to $\hat x_e
= 1.4$. By neglecting the corrections given by the metric function
$\omega$ in equations~(\ref{j-cons-iso}) and~(\ref{j-cons}), the
choice $A = \hat{R}$ leads to $x_e = \hat{x_e}$ and $\Omega_e = 0.33
\, \Omega_c$ in both coordinate systems.


\end{document}